%% file: SysControlLetter_Preprint.tex
\documentclass[a4paper,fleqn]{cas-dc}

\usepackage[numbers]{natbib}

\usepackage{graphicx}%
\usepackage{multirow}%
\usepackage{amsmath,amssymb,amsfonts}%
\usepackage{amsthm}%
\usepackage{amsxtra}
\usepackage{mathrsfs}%
\usepackage[title]{appendix}%
\usepackage{color,xcolor}
\usepackage{textcomp}%
\usepackage{manyfoot}%
\usepackage{booktabs}%
\usepackage{algorithm}%
\usepackage{algorithmicx}%
\usepackage{algpseudocode}%
\usepackage{listings}%
\usepackage{tikz}
\usepackage{subcaption}
\usetikzlibrary{arrows.meta,positioning,fit}

\theoremstyle{definition}
\newtheorem{definition}{Definition}
\newtheorem{lemma}{Lemma}

\definecolor{cbfcolor}{RGB}{245,166,35}
\definecolor{syscolor}{RGB}{52,120,188}
\definecolor{gpcolor}{RGB}{230,230,230}
\definecolor{kernelcolor}{RGB}{200,200,200}

\def\tsc#1{\csdef{#1}{\textsc{\lowercase{#1}}\xspace}}
\tsc{WGM}
\tsc{QE}
\tsc{EP}
\tsc{PMS}
\tsc{BEC}
\tsc{DE}

\begin{document}
\let\WriteBookmarks\relax
\def\floatpagepagefraction{1}
\def\textpagefraction{.001}

\shorttitle{Model-Free Based Computations of Recursive Control Barrier Function: Ultra-Local Model Approach}
\shortauthors{L. Michel, R. de Castro, J. Moyalan, I. Ebrahimi, J.-P. Barbot}

\title [mode = title]{Model-Free Based Computations of Recursive Control Barrier Function: Ultra-Local Model Approach }                     



\author[1]{L. Michel}[orcid=0000-0003-2630-7003]
\cormark[1]
\ead{loic.michel@ec-nantes.fr}
                        
\author[2]{R. de Castro}
                        
\author[2]{J. Moyalan}
                        
\author[2]{I. Ebrahimi} 
                
\author[1,3]{J.P. Barbot}


\affiliation[1]{organization={Nantes Universit\'e- \'Ecole Centrale Nantes-LS2N, UMR 6004 CNRS},
                addressline={1 rue de la Noë}, 
                city={Nantes},
                postcode={44300}, 
                country={France}}

\affiliation[2]{organization={University of California, Merced - Dep. of Mechanical Engineering},
                addressline={5200 N. Lake Road,}, 
                city={Merced},
                postcode={95343}, 
                country={CA, USA}}
                        
\affiliation[3]{organization={QUARTZ EA 7393, ENSEA},
                city={Cergy-Pontoise},
                postcode={95000}, 
                country={France}}




\begin{abstract}
Control barrier functions (CBFs) provide a systematic framework for enforcing safety constraints 
in nonlinear control systems. However, their implementation typically relies on accurate system models, 
which can limit their applicability in the presence of significant modeling uncertainties or 
unknown dynamics. This paper proposes a model-free framework for the computation of recursive control barrier 
functions based on the ultra-local model approach that leverages online estimation 
of the unknown system dynamics to construct CBF constraints. This approach does not require an explicit model of the system dynamics and 
enhances robustness with respect to disturbances and model mismatch.
The resulting control architecture enables the enforcement as well as the anticipation of 
safety constraints for systems 
with higher relative degree. 
The effectiveness of the proposed approach is illustrated on the adaptive cruise control 
benchmark.
\end{abstract}




\maketitle

\section{Introduction}

Safety-critical control problems arise in numerous engineering applications, including 
autonomous vehicles, energy systems, robotics, and aerospace systems. In such contexts, 
the controller must ensure not only performance but also strict compliance with safety 
constraints, despite model uncertainties, disturbances, and limited sensing.
Control Barrier Functions (CBFs) \cite{Ames2017CBF} \cite{Ames2019} have recently 
emerged as a powerful framework to enforce safety by guaranteeing forward invariance 
of a prescribed safe set. By constraining the admissible control inputs through Lyapunov-like 
inequalities, CBFs enable the systematic synthesis of safety filters that can be combined 
with nominal controllers while minimally degrading performance \cite{wabersich2026}.

CBFs have been successfully applied to a wide range of systems, including automotive 
applications {such as adaptive cruise control}  \cite{Xiao2010}), 
robotics \cite{Ferraguti2022,Michel2025}, and energy storage systems \cite{HAILEMICHAEL2025}. 
Many practical applications involve higher relative degree systems. 
High Order Control Barrier Functions extend the notion of CBFs to handle these cases by 
sequentially defining intermediate functions that similarly satisfy forward invariance 
conditions \cite{Nguyen2016}. In addition, cascade formulations of CBFs have been proposed 
to reduce computational burden by decomposing complex safety constraints into hierarchical 
subproblems while preserving formal guarantees \cite{Castro2022}. 

A key challenge in practical deployment, however, lies in the presence of modeling uncertainties 
and unmodeled dynamics. Classical CBF formulations typically assume accurate system models, 
which may not be available in many real-world scenarios. 
To address this limitation, recent works have explored the integration of CBFs with data-driven 
techniques \cite{Khaledi2026} and Gaussian
Process  regression as an alternative approach that
provides probabilistic guarantees of its prediction, 
 e.g. \cite{Khan2020} \cite{Castaneda2021}, and the introduction of activation functions
 to select dynamically the constraints \cite{Ong2024}.

The combination of CBFs with real-time uncertainty estimation raises several open issues 
since in practice, ensuring robustness while preserving smooth control action is critical 
for implementation on physical systems. For instance, in \cite{Ajeyemi2025}, neural networks 
have been considered to pre-calculate the CBF-based optimized controller and recently,
 in \cite{Wang2021}, estimation of the dynamics is made by an observer, and reinforcement 
 learning (RL) have been proposed \cite{Cheng2019} \cite{guerrier2024} \cite{Wijayatunga2026} 
 to take online decision with respect to RL policy in order to compute CBF-related parameters.

Other approaches inspired from model-free based control methods have been considered to estimate in 
real-time unknown dynamics using minimal prior knowledge, enabling robust control design with reduced 
modeling effort.
In particular, the model-free control, introduced by \cite{Fliess2013}, 
learns the dynamic of the system to control from an online updated approximation 
given by an ultra-local model. Such ultra-local model (ULM) is usually represented by a 
first order or second order model whose key parameter is estimated online. 
These approaches have shown promising results in automotive applications such as 
adaptive cruise control, where safety must be guaranteed despite uncertain vehicle 
dynamics and external disturbances. In particular, in \cite{PRETO2026} \cite{Chinelato2023}, sliding-mode control
and model-free control techniques have been combined to improve the online learning of the dynamics, 
while enhancing the definition of the CBF constraints. {However, such approaches typically 
rely on the instantaneous estimate of the ultra-local model without 
explicitly quantifying the uncertainty associated with the reconstruction of the dynamics. 
As a consequence, the effect of estimation errors, measurement noise and derivative approximation 
on the safety constraint is not explicitly propagated in the barrier condition. 
In addition, the ultra-local approximation is mainly used as a local representation of the dynamics, 
exploiting short-term behavior of its evolution.

These limitations motivate the development of a framework in which the uncertainty associated with the 
ultra-local model is explicitly quantified and propagated within the CBF constraint. }
In this work, we propose an {\it ultra-local model-based CBF condition} for relative-degree-two safety constraints 
under bounded model uncertainty estimated online: the system is modeled by a ULM representation, 
that allows modelling online short-term behavior of the 
    system, dynamically update the CBF constraint and enabling safety guarantees without requiring 
    a precise system model. We provide preliminary simulation results illustrating the feasibility of the approach 
    and discussing practical tuning guidelines. The simulation considers an adaptive cruise 
    controller (ACC) example, which is a common benchmark control problem within the CBF 
    community \cite{Ames2019} \cite{Agrawal2021} \cite{Xu2018}.

{The remainder of the paper is organized as follows. Section~II introduces some recalls and the problem statement. Section~III presents the proposed CBF methodology. Section ~IV, dedicated to simulations results, {highlights} the well founded of the proposed algorithms. The paper ends, in section ~V, with a conclusion and some perspectives.

\section{Some recalls and problem statement}

Consider an affine nonlinear dynamical system of relative degree\footnote{The relative degree of an input–output dynamical system is the smallest integer 
$r$ such that the $r$-th time derivative of the output depends explicitly on the input.} $r$ that reads
\begin{equation}
    \begin{aligned}
\dot{x}(t) &= f(x) + g(x)u(t)  \\
y & = h(x)
    \end{aligned}
\label{eq:system}
\end{equation}
where $x \in \mathcal{X} \subset \mathbb{R}^n$ is the system state, 
$u \in \mathcal{U} \subset \mathbb{R}^m$ the control input and $y \in \mathbb{R}$ is the output. 
The set $\mathcal{X}$ characterizes the operation domain of the system and 
$\mathcal{U}$ the input constraint set, defined as $\mathcal{U} = \{u \in \mathbb{R}^m : \underline{u} \le u \le \bar{u}\}$ where $\underline{u}$ and $\bar{u}$ represent the minimum and maximum actuation limits respectively. 

The functions $f$ and $g$ represent the nominal system dynamics, assumed to be locally Lipschitz continuous.

\subsection{Ultra-Local Modeling and Uncertainty Quantification}

Instead of relying on an explicit 
model of the system dynamics, we adopt an
ultra-local modeling framework, inspired from \cite{Fliess2013}, in which the system behavior is approximated 
locally by a low-order input–output representation that is continuously updated online, 
including a uncertainty bound $\Delta_m$ that predicts a confidence interval over a short-term period.
The local dynamics of \eqref{eq:system} is described by the SISO
ultra-local model
\begin{equation}
{y^{(\nu)}}(t) \approx F(t) + \beta u(t) + \Delta_m(t),
\label{eq:ULM_1d}
\end{equation}
where $u(t)$ denotes the control input, $\beta$ is a non-zero constant chosen
by the practitioner such as \(F(t)\) and \(\beta u(t)\) have comparable magnitudes, $y(t)$ is the ouput measurement,
$\nu \geq 1$ is the order of the derivation of $y$ and $F(t)$ represents the aggregated effect of the unknown 
system dynamics, including nonlinearities, 
disturbances, and possible modeling inaccuracies. The online estimation of $F$ assumes firstly that the output $y$ is available from measurements and
continuously compensates for the discrepancy between the simplified ultra-local 
representation and the actual system behavior. {In fact, it is well known, since the seminal paper 
\cite{HAUSER1992} about PVTOL and see \cite{riachy:inria-00463605} for a model-free 
control application, that for the purpose of zero dynamics stability, it is necessary to 
consider $\nu$ greater than the relative degree. For the sake of simplicity, we assume that the zero dynamics of the original system is input-to-state stability (ISS) \cite{SONTAG1995}\footnote{For the stability proof in this context and the link with PID controllers, see \cite{praly:hal-00531004}. To have a well conditioned control and good controlability, $\beta$ should be chosen such as $|\beta| > \beta_{min}$ where $\beta_{min}$ is a positive minimal value.}.}

The term $F(t)$ is estimated online from measurements using numerical
differentiation techniques, leading to an estimate $\hat{F}(t)$. The estimation
process is affected by measurement noise and approximation errors, which result
in an uncertainty on the reconstructed dynamics.
\subsection{Background on Control Barrier Functions}

In this section, the basic definitions and concepts of control barrier functions are introduced \cite{Castro2022}, for which 
we focuse on the form of \textit{zeroing control barrier functions (ZCBFs)}.
Before presenting the definition of the ZCBF, let us introduce the motivation.
Assume that we have a constraint
\begin{equation}
\exists \, u \in \mathcal{U} \quad \text{s.t.} \quad {c}(y,u) \ge 0 \quad \forall y \in \mathcal{D}
\label{c_constraint}
\end{equation}
\noindent
where $\mathcal{D}$ encloses the safe set $\mathcal{C}$, i.e. $\mathcal{C} \subset \mathcal{D}$.
If this condition can be enforced and $c(y_0) \ge 0$ then the value of the boundary function $c$ is monotonic increasing 
and it will never reach negative values (i.e. $y$ will never go outside $\mathcal{C}$); hence $\mathcal{C}$ will be invariant.
However, this condition requires $c$ to be always non-negative, which might be very restrictive. 
The ZCBF avoids such conservatism by allowing $\dot{c}$ to reach negative values, especially when the state moves 
toward the interior of $\mathcal{C}$. Accordingly, the ZCBF replaces \eqref{c_constraint} with the following relaxed constraint
\begin{equation}
\exists \, u \in \mathcal{U} \quad \text{s.t.} \quad \dot{c}(y,u) \ge -\alpha(c(y)), 
\quad \forall y \in \mathcal{D}
\label{c_constraint_alpha}
\end{equation}
\noindent
where $\alpha$ is an extended class-$\mathcal{K}$ function. 
In the case of an undesirable situation for which larger perturbations than expected might occur, the boundary function $c$ is negative, and consequently \eqref{c_constraint_alpha} guarantees $\dot{c} > 0$. Indeed \eqref{c_constraint_alpha} increases the value of $c$ and attracts the state 
to the interior of $\mathcal{C}$. It can be also verified that $\dot{c} = 0$ at the boundary of $\mathcal{C}$, which prevents the boundary function $c$} from becoming negative at the boundary of the safety set, 
which simultaneously avoids the system state from leaving $\mathcal{C}$.

The ZCBF is formally defined as follows.

\begin{definition}\cite{Ames2017CBF}
Consider the set $\mathcal{C}$ defined by (3). 
The boundary function $c(y)$ is a zeroing control barrier function (ZCBF) defined on set $\mathcal{D}$, 
with $\mathcal{C} \subset \mathcal{D} \subset \mathcal{X}$, if there is an extended class-$\mathcal{K}$ function 
$\alpha$ such that
\begin{equation}
\sup_{u \in \mathcal{U}} 
\left[ L_f c(y) + L_g c(y)u + \alpha(c(y)) \right] \ge 0,
\quad \forall y \in \mathcal{D}
\label{eq:def_CBF}
\end{equation}
\noindent
The feasible input set capable of enforcing (6) is represented as
\begin{equation}
\mathcal{U}_c(y) =
\{u \in \mathcal{U} : L_f c(y) + L_g c(y)u + \alpha(c(y)) \ge 0 \}
\end{equation}
\noindent
where $L_f c(y) = \frac{\partial c(y)}{\partial y} f(y)$ is the Lie derivative of $c(y)$ 
along the vector field $f(y)$. 
\end{definition}
\noindent
Forward invariance of $\mathcal{C}$ follows from the next lemma.
 \begin{lemma}\cite{Ames2017CBF}
Consider the safe set $\mathcal{C}$ and the ZCBF $c(y)$ defined on $\mathcal{D}$. 
Any Lipschitz continuous controller $u$ such that $u \in \mathcal{U}_c(y) \neq \emptyset$, 
$\forall y \in \mathcal{D}$ for the system (1) renders the set $\mathcal{C}$ forward invariant, 
i.e., for all $y(0) \in \mathcal{C}$, $y(t) \in \mathcal{C}$, $t \ge 0$.
\end{lemma}
\noindent
As depicted in Fig.~\ref{archi_CBF}, ZCBFs can be integrated into control loops as a means to enforce safety. 
This integration is often achieved through numerical optimization
\begin{equation}
u^*(y) = \arg\min_{u(y) \in \mathcal{U}_c(y)} \|u - u_{nom}(t)\|.
\label{eq:u_optim}
\end{equation}
\noindent
For the sake of simplicity of notation, $u^*$ depends on $y$ to highlight the safety considerations.
Consider a nominal control $u_{nom}(t)$ that is designed as the feedback control policy for 
system \eqref{eq:system} designed by the practitioner sing for instance physics-based, 
data-driven or model-free controllers.
The safety of the system cannot be guaranteed with such a
given nominal control policy.
\subsection{Problem Statement}
\label{problem_statement}
We are interested in designing control input $u^*(y)$ \eqref{eq:u_optim} that renders \eqref{eq:system} safe while encoding 
the unmodelled dynamics of \eqref{eq:system} in the design of $c(y)$ through an ultra-local model representation \eqref{eq:ULM_1d}.
To this end, we assume the existence of a nominal controller $u_{nom}(t)$  to achieve 
the desired performance objective. However, this nominal control does not necessarily guarantee 
safety with respect to the constraint $c(y) \geq 0$.
CBFs are therefore employed to modify the nominal control input through an ultra-local model-based
safety filter, while minimally deviating from $u_{nom}(t)$.
\section{Ultra Local Model-CBF}

{Starting from the point of view of the practitioner who designed the nominal control, our goal is to design a control which verify the safety constraints. 
For this purpose, the ultra-local method removes any dependencies with respect to the modeling and the considered nominal control.}
This section presents the proposed {ULM-CBF} framework. 
The method relies on the estimation of the uncertainty bound $\hat{\Delta}_m(t)$.  Hence the ultra-local model \eqref{eq:ULM_1d} can be rewritten
\begin{equation}
\hat{y}^{(\nu)}(t) \approx \hat{F}(t) + \hat{\beta} u(t) + \delta
\label{eq:ULM_Delta_m}
\end{equation}

\noindent
with $\delta \in [-\hat{\Delta}_m(t), \, \hat{\Delta}_m(t)]$. The quantity $\hat{\Delta}_m(t)$ represents a worst-case bound on the estimation
error of the lumped dynamics. In practice, it depends especially on the quality
of the derivative estimation and on the level of measurement noise. 
To numerically estimate the $\nu$-th derivative\footnote{In the following, the order $\nu$ of the ultra-local model \eqref{eq:ULM_1d} is selected to match the relative degree $r$ of the safety constraint $c(y)$.} of $y$, homogeneous semi-implicit differentiators 
have been used (see \cite{Michel_al2021, CEP_Rasool, Springer_loic}) since they gives remarkable properties of filtering 
and allow higher-order derivatives. 
The bound $\hat{\Delta}_m(t)$ allows the safety constraints to be
formulated in a "robust" manner, which remains valid despite
estimation errors in the ultra-local model. In particular, the barrier condition can be
modified to account for the worst-case deviation between the true dynamics and
their estimated counterpart. Two strategies are proposed: the strategy 1 focuses on a gradient-based ULM-CBF that calculates 
the resulting safe control and the strategy 2 presents a gridding approach enabling a decision-making to calculate the safe control.
Figure \ref{archi_CBF} illustrates the proposed CBF architecture including the ultra-local model and the $\hat{\Delta}_m$ update.

\input{Figures/Architecture_scheme_v1.tex}
\subsection{Strategy 1:Gradient-Based ULM-CBF}
\subsubsection{Estimation of $(F, \beta)$}
\label{lem:ULM_estimation}
\noindent
Consider the ULM {with $\nu=1$}
\begin{equation}
    \frac{d \hat{y}}{dt}(t) = \hat{F}(t) + \hat{\beta}(t) u (t) + \delta
\end{equation}
\noindent
with $\delta \in [-\hat{\Delta}_m(t), \, \hat{\Delta}_m(t)]$. In order to implement the strategy in the 
software in the loop, let us consider the following discrete-time values $\dot{y}_k = \frac{d y}{dt}(k T_s), \hat{F}_k = \hat{F}(k T_s), $ etc, where $T_s$ is the sampling time.
The associated discrete-time equation of the ULM is defined as
\begin{equation}
\dot y_k = \hat{F}_k + \hat{\beta}_k u_k + \delta_k,
\end{equation}
\noindent
with $\delta \in [-\hat{\Delta}_m(t), \, \hat{\Delta}_m(t)]$. To achieve a robust online reconstruction of the system behavior, we adopt a recursive estimation framework based on a 
two-time-scale stochastic gradient descent. The ultra-local model (ULM) can be rewritten as an optimization problem such as
the estimation objective is to minimize the instantaneous quadratic cost function $J_k(\theta)$
\begin{equation}
    J_k(\theta) = \frac{1}{2} \left( \hat{\dot{y}}_k - \phi_k \theta \right)^2.
    \label{Jcostcriteria}
\end{equation}
\noindent
The right part of the ULM $\hat{\dot{y}}$ is the estimated time derivative of $y$ and the left part of the ULM is written as a regressor formulation
$\phi_k \theta_k$  where $\theta_k = [\hat{F}_k, \hat{\beta}_k]^\top$ is the vector of unknown parameters and $\phi_k = [1, u_k]$ is the regressor vector sufficiently variable.
The corresponding stochastic gradient descent (see e.g. \cite{Lennart}) approach reads
\begin{equation}
    \theta_{k+1} = \theta_k - \lambda_\theta \nabla_\theta J_k(\theta_k) = \theta_k + \lambda_\theta \phi_k \left( \hat{\dot{y}}_k - \phi_k \theta_k \right)
    \label{gradient_ULM}
\end{equation}
where $\lambda_\theta = \text{diag}([\lambda_F, \lambda_\beta])$ and $\lambda_F >0$, $\lambda_\beta > 0$ 
represents the matrix of adaptation gains. 
This formulation ensures that both parameters are updated consistently to minimize the local 
reconstruction error of the system dynamics. In particular, 
In this framework, setting $\lambda_F \neq \lambda_\beta$ allows for a multi-time-scale 
adaptation strategy: $\hat{\beta}$  is updated to capture rapid variations in the dynamics, 
while the lumped term $\hat{F}$ is adapted more conservatively to identify the structural input sensitivity 
of the system.
\subsubsection{The $\hat{\Delta}_m$ computation}
The evolution of the ultra-local term $\hat{F}$ reflects the local variations
of the system dynamics along the trajectory. Indeed, within the ultra-local
modeling framework, variations of $\hat{F}$ capture the local changes of
the underlying dynamics $f(x)$ projected onto the input–output representation.
From \eqref{Jcostcriteria}, define $\delta_k$ the residue of the $F$ estimation such as
$\delta_k = \hat{\dot{y}}_k - \phi_k \theta$.
\paragraph{Mean and Variance of $\delta$}

The Welford’s algorithm \cite{welford1962} is a numerically stable algorithm for 
computing the running mean and variance of a sequence of observations. 
This formulation avoids the loss of precision associated with classical variance formulas and is 
therefore well suited for online data processing.
We calculate a statistical average based on a finite amount of data, based on a recursive formulation of the original algorithm.
To do this, we use a First-In First-Out (FIFO) shift register of finite length $n$ as a moving sliding window slot from which samples are collected.

Given a sequence $\delta_{(k-n)h}, \dots, \delta_{kh}$, at time $t=kh$, this sequence become 
at time $t=(k+1) h$: $\delta_{(k-n+1)h},$ $\delta_{(k-n+2)h}, \dots, \delta_{(k+1)h}$, 
the Welford update for the mean $\mu_{(k+1)h}$ and the centered second moment $M_{(k+1)h}$ is
\begin{align}
\mu_{(k+1)h} &= \mu_{kh} + \frac{\delta_{(k+1)h} - \delta_{(k-n+1)h}}{n}, \nonumber  \\
M_{(k+1)h} &= M_{kh} + (\delta_{(k+1)h} - \mu_{kh})(\delta_{(k+1)h} - \mu_{(k+1)h})\nonumber \\
&-(\delta_{(k-n+1)h} - \mu_{kh})(\delta_{(k-n+1)h} - \mu_{(k+1)h}), \label{eq:M}
\end{align}
with $\delta_{lh}=0$  and $\mu_{lh}=0$ for $l=0$. Equation \eqref{eq:M} gives the square variance 
estimate $\displaystyle{\sigma_{kh}^2 = \frac{M_{kh}}{n}}$ and  $\sigma_k$ is the computed standard deviation at the instant $k$. In practice, $n$ can be chosen as low as possible in order to avoid estimation delays and a quick convergence on $\sigma^2$.


We construct $\Delta_m$ using a deterministic–statistical strategy based on the online estimate 
$\hat{F}$. To compute the uncertainty bound $\hat{\Delta}_m$, three complementary formulations are proposed. Method A adopts a spatial-geometric approach where the uncertainty envelope is scaled according to the dynamics's proximity to the safety boundary. Method B combines this spatial-geometric penalty layer with the Welford statistical variance to protect against transient mismatch spikes. Method C relies solely on a statistical variance threshold.
\\
\\
\noindent
{\bf Method A:} The $\Delta_m$ bound is defined by
\begin{equation}
\hat{\Delta}_{m_k} = \varepsilon_k^{MAX}  \frac{1}{1 + |y_k - y_{min}|}
\end{equation}
\noindent
where $\varepsilon_k^{MAX}$ holds the maximum value of $| {\mu_k} | $ such as
\begin{equation}
\varepsilon_k^{MAX} = \sup(| \mu_k | ,  \, \mathcal{D}(y_k)   \varepsilon_{k-1}^{MAX}, \varepsilon_p^{MAX} ),
\end{equation}

\noindent
and $\mathcal{D}(y_k) = \frac{1}{1 + \varepsilon_p  | y_k - y_{min}|}$ defines a forgetting factor that depends on the distance between $y_k$ and  $y_{min}$. 
\noindent
The {\it geometric penalty}  $\varepsilon_p$ is chosen such that $\mathcal{D}(y_k) \lesssim 1$ for $y_k \approx d_{min}$.
\\
\noindent
{\bf Method B:} The $\Delta_m$ bound is defined by

\begin{equation}
\hat{\Delta}_{m_k} = \varepsilon_k^{MAX}  \frac{1}{1 + |y_k - y_{min}|}
\end{equation}

\noindent
where $\varepsilon_k^{MAX}$ holds the maximum value of $| {\mu_k} | + | \sigma_k | $ such as
\begin{equation}
\varepsilon_k^{MAX} = \sup(| {\mu_k} | + | \sigma_k | ,  \, \mathcal{D}(y_k)   \varepsilon_{k-1}^{MAX}, \varepsilon_p^{MAX} ),
\end{equation}

\noindent
and $\mathcal{D}(y_k) $ defines a forgetting factor as previously described.
\\

Hence, in methods A and B, the maximum value $\varepsilon_k^{MAX}$ is hold while $y_k$ remains close to $y_{min}$ in case of very critical safety operation.
The lowest bounds $\varepsilon_p^{MAX} , \varepsilon_{\sigma}$ define minimal values of respectively $\varepsilon_k^{MAX}$ and $\sigma_k$ (avoid cancellation if the noise if very small). 
\\
\noindent
{\bf Method C:}
The $\Delta_m$ bound is defined by
\begin{equation}
\hat{\Delta}_{m_k} = | {\mu_k} | + \kappa |\sigma_k|
\end{equation}
where $\kappa$ is a variance threshold parameter. 
Note that methods A and B include a "spatial geometry layer", whereas method C is based only on statistical propagation of $\delta$.
\\
The $\Delta_m$  bound combines three complementary indicators computed over the Welford sliding window (Fig. \ref{Welford_scheme}): 
\\
(i) the average magnitude of the residual model mismatch $\delta$ which ensures that the uncertainty bound reflects the current quality of the ULM identification 
\\
(ii) the empirical standard deviation of $\delta$
which provides a deterministic safety envelope that protects the system against transient spikes and sudden 
disturbances that purely statistical moments might temporarily smooth out and, in the methods A and B,  
\\
(iii) a non-linear weighting 
factor based on the system's proximity to the safety boundary. By amplifying $\Delta_m$ as the state $y$ 
approaches $y_{min}$, the framework dynamically shifts the virtual barrier 
to be more restrictive in critical regions. Method C requires less computational effort than method A and B.

\input{Figures/Welford_scheme_v1.tex}

Combining the worst-case, first-order (structured mismatch) and second-order statistical (stochastic uncertainty) information 
of uncertainty leads to the bound $\hat{\Delta}_m$, which captures both instantaneous 
excursions and statistical fluctuations of the unknown dynamics and which represents 
an online estimate of a {\it confidence envelope} on the predicted ultra-local dynamics. 
It enables the formulation of a robust control barrier function constraint without requiring 
any explicit model information, while remaining compatible with real-time implementation.

This envelope is propagated within the barrier constraint in order to
preserve safety despite estimation errors and local variations of the
system dynamics.
Figure \ref{ULM_Deltam} depicts the propagation of the uncertainty envelope $\hat{\Delta}_m$ associated
with the ULM.
\input{Figures/ULM_Delta_m.tex}
\subsubsection{Robust CBF with Model Uncertainty}

Let the safety constraint be defined by a linearly as
\begin{equation}
c(y) = y - y_{min} \ge 0.
\end{equation}
where $y_{min}$ is the minimum safety distance. 
\paragraph{Robust CBF for systems with relative degree 1}

According to \eqref{eq:def_CBF}, \eqref{eq:ULM_Delta_m}, the CBF condition reads
\begin{equation}
\dot{c}(y) = \frac{\partial  c}{\partial y} ( F + \beta u ) \geq - \alpha(c(y)),
\label{eq:stand_CBF}
\end{equation}

\noindent
and setting $\alpha_c(c) = k_c c$, 
the standard CBF condition \eqref{eq:stand_CBF} becomes 
\begin{equation}
\hat{F} + \beta u \ge -k_c (y - y_{\min}).
\label{cbf_standard}
\end{equation}

\noindent
The corresponding { CBF } accounting for model uncertainty $\hat{\Delta}_m$ reads
\begin{equation}
\mathcal{H}_{r = 1} := \hat{F} + \beta u + k_c \bigl(y - y_{\min}\bigr) - \hat{\Delta}_m \ge 0
\end{equation}

\noindent
where $k_c > 0$ is a design parameter. This inequality can be rewritten as
\begin{equation}
\hat{F} + \beta u \ge -k_c\left(y - y_{\min}^{\mathrm{shifted}} \right)
\label{cbf_deltam}
\end{equation}
where $y_{\min}^{\mathrm{shifted}} = y_{\min} + \frac{\hat{\Delta}_m}{k_c}$. Hence, the robust CBF condition \eqref{cbf_deltam} is equivalent to a standard CBF constraint \eqref{cbf_standard} applied to a \emph{virtual safety boundary} $(y_{\min}^{\mathrm{shifted}})$, leading to earlier activation of the CBF constraint and improved anticipation of safety violations. 

\noindent
The gain $k_c$ adjusts both the corrective action of the CBF dynamics and the magnitude of the virtual barrier shift induced by uncertainty $\hat{\Delta}_m$.

\noindent
The optimization problem 
$P_{CBF-1}(\hat{\beta}, \hat{F},\Delta_m^{},u_{nom})$ reads as

\begin{equation}
\begin{aligned}
P_{CBF-1}: 
\min_{u \in \mathcal U} \quad & (u-u_{\mathrm{nominal}})^2 \\
\text{s.t.} \quad
&
\hat{F} + \hat{\beta} u + k_c \bigl(y - y_{\min}^{\mathrm{shifted}}\bigr) \ge 0 \\
\end{aligned}
\end{equation}

\paragraph{Robust CBF for systems with relative degree 2}

Considering the relative degree 2, from \cite{Castro2022}, the general definition of the CBF \eqref{eq:def_CBF} becomes
\begin{equation*}
\small{\sup_{u \in \mathcal{U}}
\left[ L_f^2 c(y) + L_g L_f c(y)u + \begin{pmatrix} k_{0} & k_{1} \end{pmatrix} \begin{pmatrix} c(y) \\  \dot{c}(y) \end{pmatrix} \right] \ge 0, \forall y \in \mathcal{D}}
\label{eq:def_CBF_2}
\end{equation*}
\noindent
where $k_0$ and $k_1$ are the dynamic gains of the CBF. By identifying the terms $L_f^2 c(y)$ and $L_g L_f c(y)$ with the ULM \eqref{eq:ULM_1d}, we have
\begin{equation}
\ddot c(t = kh) = \left. L_f^2 c(y) \right|_{x(kh)} + \left. L_g L_f c(y) u(kh) \right|_{x(kh)}
\end{equation}
\noindent
Hence, the ULM \eqref{eq:ULM_Delta_m} at the second order reads
\begin{equation}
\ddot c_k = \hat{F}_{2k} + \hat{\beta}_{2 k} u_k + \delta_k
\label{eq:ulm2} 
\end{equation} 
\noindent
where $\delta \in [-\hat{\Delta}_m(t), \, \hat{\Delta}_m(t)]$ and $\ddot c_k = \ddot y_k$. The { CBF } accounting for model uncertainty $\hat{\Delta}_m$ reads
\begin{equation}
\mathcal{H}_{r = 2} := \hat{F}_2 + \beta u + k_1 \dot y + k_0 (y - y_{\min}^{\mathrm{shifted}}) > 0
\end{equation}
where $\hat{F}_2$ is computed from the second order time derivative\footnote{The computation of high order time-derivative constitutes a practical challenge due to the presence of noise. However, as stated before, the high order time-derivatives can be computed using semi-implicit differentiators (see \cite{Michel_al2021, CEP_Rasool, Springer_loic}) since it is assumed that the samples $\{\ddot c_k, u_k\}$ are accessible.} of $y$ and $y_{\min}^{\mathrm{shifted}} = y_{\min} + \frac{{\hat{\Delta}}_m}{k_0}$.

\noindent
The optimization problem 
$P_{CBF-2}(\hat{\beta}, \hat{F},\Delta_m^{},u_{nom})$ reads as

\begin{equation}
\begin{aligned}
P_{CBF-2}: 
\min_{u \in \mathcal U} \quad & (u-u_{\mathrm{nominal}})^2 \\
\text{s.t.} \quad
&
\hat{F}_2 + \hat{\beta} u + k_1 \dot y + k_0 (y - y_{\min}^{\mathrm{shifted}}) \geq 0 \\
\end{aligned}
\end{equation}

{\subsection{Strategy 2: Grid-ULM-based CBF}}

In the previous section, we already presented a strategy to estimate online $\beta$. This section provides a 
second approach to estimate $\beta$ based on gridding in which several ultra-local models are evaluated in parallel 
using a discretized set of $\beta$ values.
The resulting uncertainty envelope depends on the "quality" of the
ultra-local approximation, which in turn is influenced by the choice of the scaling parameter $\beta$. 

A discretized set of candidate parameters $\mathcal B = \left\{ \beta_1,\dots,\right. $ $\left.\beta_j,\dots,\beta_M \right\}$, that constitutes a grid of $M$ points, is 
introduced. The algorithm~\eqref{Algo1} describes the proposed $\beta$-gridding procedure.
At each instant $k$, a nominal controller allows calculating a nominal control $u_{nom}$ that would give 
an expected trajectory of the controlled $y$ (see Section \ref{problem_statement}).
For each candidate ultra-local model, associated to the element $\beta_j$ of $\mathcal B$, a robust CBF-based
quadratic program is solved. The solution of this QP program provides a control candidate $u_{j,k}^{cbf}$ 
for $j = 1...M$ which is consistent with the corresponding estimated ultra-local dynamics $\hat{F}_{j,k}$ 
and the associated
uncertainty envelope $\hat{\Delta}_{m \, {j,k}}$. 
These candidates $u_{j,k}^{cbf}$ are then evaluated through a selection
criterion in order to determine the most suitable index $j_{\mathrm{safe}, k}$ (among the $\beta_j$ parameters 
of $\mathcal B$) that gives the safest control $u_k^* = u_{j_{\mathrm{safe}, k},k}^{cbf}$ that is applied afterwards to the system \eqref{eq:system}. The proposed criteria include 
the energy of the control ($J_1$), 
the energy of the $\Delta_m$ bound ($J_2$), the variation of the control by comparing all candidates 
$u_{i,k}^{cbf}, j = 1...4$ with the previous safe control $u_{k-1}^*$ ($J_3$), and the variation 
of the selected index by holding the previous $\beta^{*}_{k-1}$ index if no substantial changes 
have been deduced from the model selection procedure ($J_4$).
\\
\subsubsection*{Optimization of the safe control}
The difference between strategies 1 and 2 is that at each instant $k$, the optimization problem $P_{CBF}$ is computed only one 
time in strategy 1 (using the estimated $\beta$ through the gradient descent \eqref{gradient_ULM}) 
whereas is it computed $M$ times in strategy 2 using the $\beta$-gridding.
\noindent

\begin{algorithm}[t]
\caption{$\beta$-gridding selection for ULM-based CBF}
\begin{algorithmic}[1]
\Require Assume a set of candidate parameters $\mathcal{B}$
\Require $\gamma_j, j =1\cdots 4$ are weights coefficients.
\State $u_0^{*} = 0$
\State $u_0^{nom} = 0$

\State \textbf{Step 1: Nominal controller}
    \State Compute $u_k^{nom}$

\State \textbf{Step 2: Estimation of $F_k$}
\For{$j=1,\dots,M$}
    \State Compute uncertainty bound $\hat{\Delta}_{m \, {j,k}}$
    \State Estimate $\hat{F}_{j,k}$ from $u_{k}^{*}$
\EndFor

\State \textbf{Step 3: CBF filtering}
\For{$j=1,\dots,M$}
    \State $u_{j,k}^{cbf} = \text{solution}\big(P_{CBF}(\beta_j,\hat{F}_{j,k},\hat{\Delta}_{m \, {j,k}},u_{k, nom})\big)$
\EndFor

\State \textbf{Step 4: Model selection}

\For{$j=1,\dots,M$}
\State Compute four cost functions for each $\beta_j$ such as:
\State $J_{j,1} = (u_{j,k}^{cbf})^2$
\State $J_{j,2} = \Delta m^2_{j,k}$
\State $J_{j,3} = (u_{j,k}^{cbf} - u^*_{k-1})^2$
\State $J_{j,4} = (\beta_j - \beta^*_{k-1})^2$
\EndFor
\State Select the index $j_k$ minimizing:
\State $j_{{\mathrm{safe}}, k} = \arg\min_j \left(\sum_{l=1}^4 \gamma_l J_{i,l} \right)$

\State \textbf{Step 5: Control update}
\State $u^{*}_k = u_{j_{\mathrm{safe}, k},k}^{cbf}$
\State $\beta^{*}_{k} = \beta_{j_{\mathrm{safe}, k}}$

    \State Apply $u_k^{*}$ to the system \eqref{eq:system}

\end{algorithmic}
\label{Algo1}
\end{algorithm}
\section{Simulation results}

Two examples\footnote{Simulation code can be found in the repository:
\\
$<$\texttt{https://github.com/LoicMichelControl/CBF$\_$ULM}$>$.} illustrate our proposed approach: the first example depicts a simple first order system for which the
strategy 1 is compared with the strategy 2. The second example illustrates the gridding strategy applied to the active cruise control
benchmark.
{\subsection{Illustrative first-order example}}

Consider the following model
$$\dot{y}(t) = - (1.13 + \sin(0.1 t))\,y(t) + (1.36 + \cos(0.1 t)) u(t)$$
that is controlled, for example, using the Model-Free Control \cite{Fliess2013}, for which $u_{nom}(0) = 0$. 
Set $x(0) = 1$ and $y^*(0) = 1$ and the following CBF parameters are used: $y_{min} = 0.4$ and $k_c = 8$. 
The gradient-based algorithm \eqref{gradient_ULM} is initialized at $\beta_0 = 0$ and $\hat{F} = 0$.





\vspace{0.2cm}
{{\it The goal is to illustrate the evolution of the controlled tracking with respect to a deliberate violation of the safety distance. }}
\vspace{0.2cm}

The illustrative examples show the tracking of a reference over 800 s that decreases below the minimum safety distance $y_{min}$ (between 200 s and 600 s).

To evaluate the practical effectiveness of the robust safety filter, a comprehensive 
parametric study has been conducted to compare the $\Delta_m$ formulations. The system's 
performance is analyzed through  multi-objective criteria: maximizing the global safety margin\footnote{The global safety margin is reflected by the average distance to 
the constraint over the $[200, 600]\,\text{s}$ window.}, ensuring strict deterministic 
safety at the critical transient peak ($\min(y) > 0$), and minimizing output control variations.
In the method 1, the optimization parameters $\lambda_F = \lambda_{\beta} = 0.1$ have been chosen\footnote{
The selection of the adaptation gains $\lambda_F$ and $\lambda_\beta$ reflects a key compromise between reactivity and stability. Small adaptation gains yield overly sluggish estimator dynamics, resulting in a persistent residual tracking error during transients. Conversely, excessively high gains destabilize the gradient estimator and induce severe safety violations. The setting $\lambda_F = \lambda_\beta = 0.1$ thus provides a good trade-off. The complete table is available in the GitHub repository.}
and in the gridding 
algorithm of the method 2 (Algorithm \ref{Algo1}), the set $\mathcal{B} = \{0.6, \, 0.8, \, 1.0, \,  1.2, \, 1.4 \}$ has been selected.

\paragraph{Parametric study of $k_c$}

In this study, the geometric penalty thickness is fixed at a nominal value ($\epsilon_p = 10^{-2}$), while the barrier stiffness is swept through $k_c \in \{0.1, 2.0, 5.0, 10.0\}$. The results are presented in Tab. \ref{tab_1} (bold indicates constraint violations / crash in Method 1 and safe restoration in Method 2).

\begin{table}[h!]
\centering
\caption{Parametric $k_c$ sweep at fixed geometric penalty $\epsilon_p = 0.01$ (Nominal  effort $\Delta u_{\text{nom}} = 99.64$).}
\label{tab:kc_sweep_complete}
\fontfamily{ptm}\selectfont
\scriptsize
\setlength{\tabcolsep}{2.5pt}
\begin{tabular}{cc r ccc}
\toprule
\textbf{Method} & \textbf{Case} & \multicolumn{1}{c}{$k_c$} & $\min(y)$ & average ${y}$ & $\Delta u$ \\
\midrule
& & 0.1 & +0.09 & +0.11 & 22.1 \\
& \textbf{A} & 2.0 & +0.018 & +0.037 & 22.1 \\
& & 5 & +0.013 & +0.027 & 22.1 \\
& & 10 & +0.01 & +0.023 & 22.1 \\
& & 100 & +0,6122 & +6,8149 & 22.1 \\
\cline{2-6}
\textbf{Method 1} & & 0.1 & +0.1076 & +0.1352 & 22.12 \\
\textbf{(No Grid)} & \textbf{B} & 2 & +0.0254 & +0.0772 & 22.12 \\
& & 5 & \textbf{-0.302} & \textbf{-0.22} & 22.12 \\
& & 10 & \textbf{-0.304} & \textbf{-0.17} & 22.12 \\
& & 100 & \textbf{-0.300} & \textbf{-0.2952} & 22.12 \\
\cline{2-6}
& & 0.1 & -0.0009 & +0.0210 & 22.12 \\
& \textbf{C} & 2.0 & -0.0002 & +0.0021 & 22.12 \\
& & 5 & -0.0001 & +0.0016 & 22.12 \\
& & 10 & \textbf{-0.0000} & \textbf{+0.0015} & \textbf{22.12} \\ 
& & 100 & \textbf{-0.3} & \textbf{-0.29} & 22.12 \\
\midrule
\midrule
& & 0.1 & +0.0902 & +0.1497 & 36.05 \\
& \textbf{A} & 2.0 & +0.0176 & +0.0302 & 36.05 \\
& & 5 & +0.0099 & +0.0251 & 36.05 \\
& & 10 & +0.0067 & +0.0229 & 36.05 \\
& & 100 & +0,0036 & +0,0224 & 36.05 \\
\cline{2-6}
\textbf{Method 2} & & 0.1 & +0.1037 & +0.1313 & 36.05 \\
\textbf{(Gridding)} & \textbf{B} & 2.0 & +0.0195 & +0.0713 & 36.05 \\
& & 5 & \textbf{+0.0129} & \textbf{+0.0372} & \textbf{36.05} \\
& & 10 & \textbf{+0.0079} & \textbf{+0.0412} & \textbf{36.05} \\
& & 100 & \textbf{+0,0036} & \textbf{+0,0214} & \textbf{36.05} \\
\cline{2-6}
& & 0.1 & -0.0003 & +0.0182 & 36.05 \\
& \textbf{C} & 2 & -0.0000 & +0.0017 & 36.05 \\
& & 5 & -0.0000 & +0.0015 & 36.05 \\
& & 10 & \textbf{-0.0001} & \textbf{+0.0014} & \textbf{36.05} \\
& & 100 & \textbf{-0,0004} & \textbf{+0,0017} & \textbf{36.05} \\
\bottomrule
\end{tabular}
\label{tab_1}
\end{table}

\begin{figure}[pos=t]
  \centering
  \begin{subfigure}{\linewidth}
    \centering
    \includegraphics[width=0.9\linewidth]{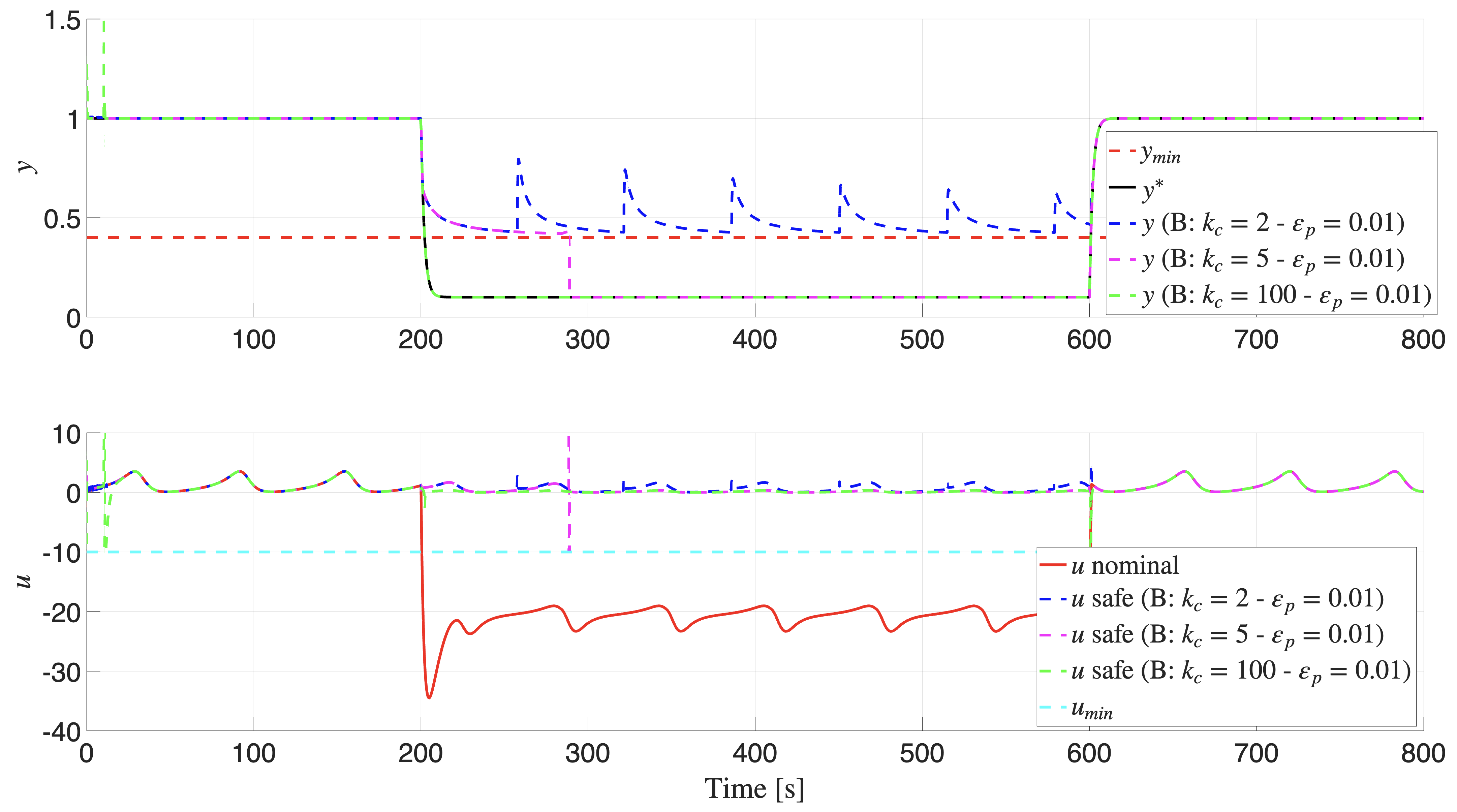}
    \caption{Method 1}
    \label{sweeping_kc_a}
  \end{subfigure}
  \vspace{0.5em} 
  \begin{subfigure}{\linewidth}
    \centering
    \includegraphics[width=0.9\linewidth]{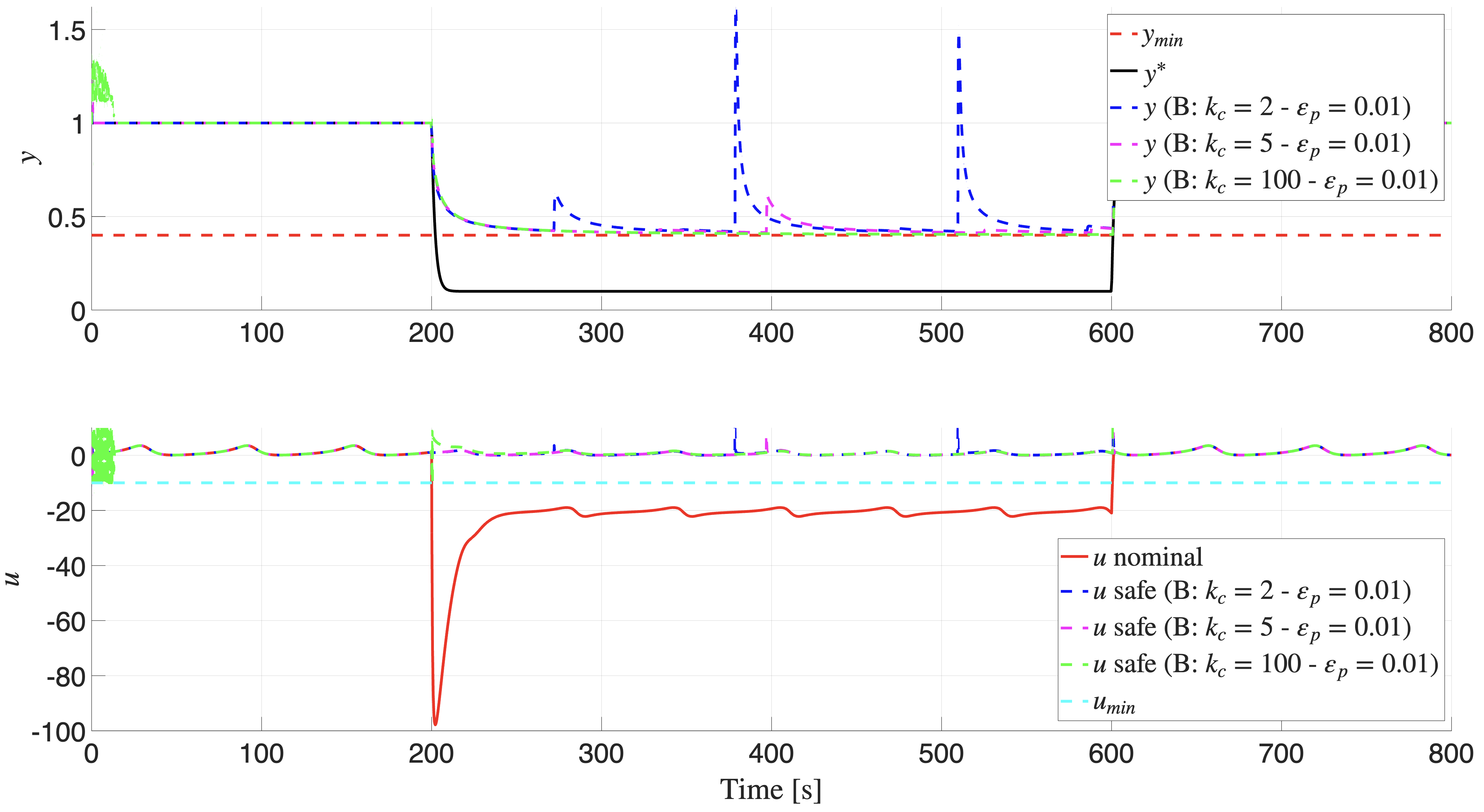}
    \caption{Method 2}  
    \label{sweeping_kc_b}
  \end{subfigure}
  \caption{First-order system - reference tracking under $k_c$ sweeping for the method B  (using $\lambda_F = \lambda_{\beta} = 0.1$ and $\varepsilon_p = 0.01$).}
  \label{fig:ensemble}
\end{figure}

\begin{figure}[pos=t]
         \centering
         \includegraphics[width=0.5\textwidth]{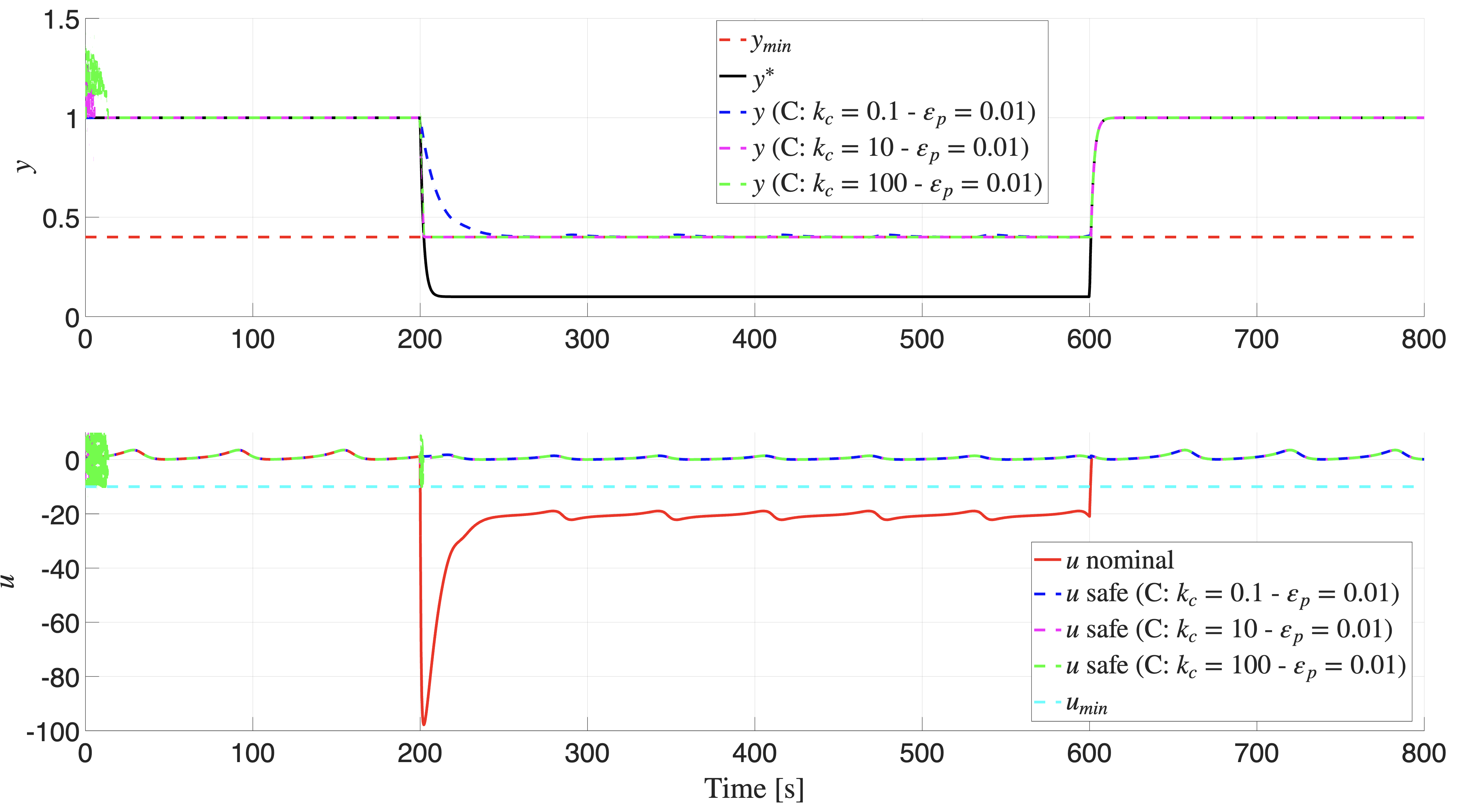}
         \caption{Method 2 for a first-order system - reference tracking under $k_c$ sweeping for the method C  ($\lambda_F = \lambda_{\beta} = 0.1$ and $\varepsilon_p = 0.01$).}
         \label{sweeping_fig_1c}
        \end{figure}

\begin{figure}[pos=t]
  \centering
  \begin{subfigure}{\linewidth}
    \centering
    \includegraphics[width=0.9\linewidth]{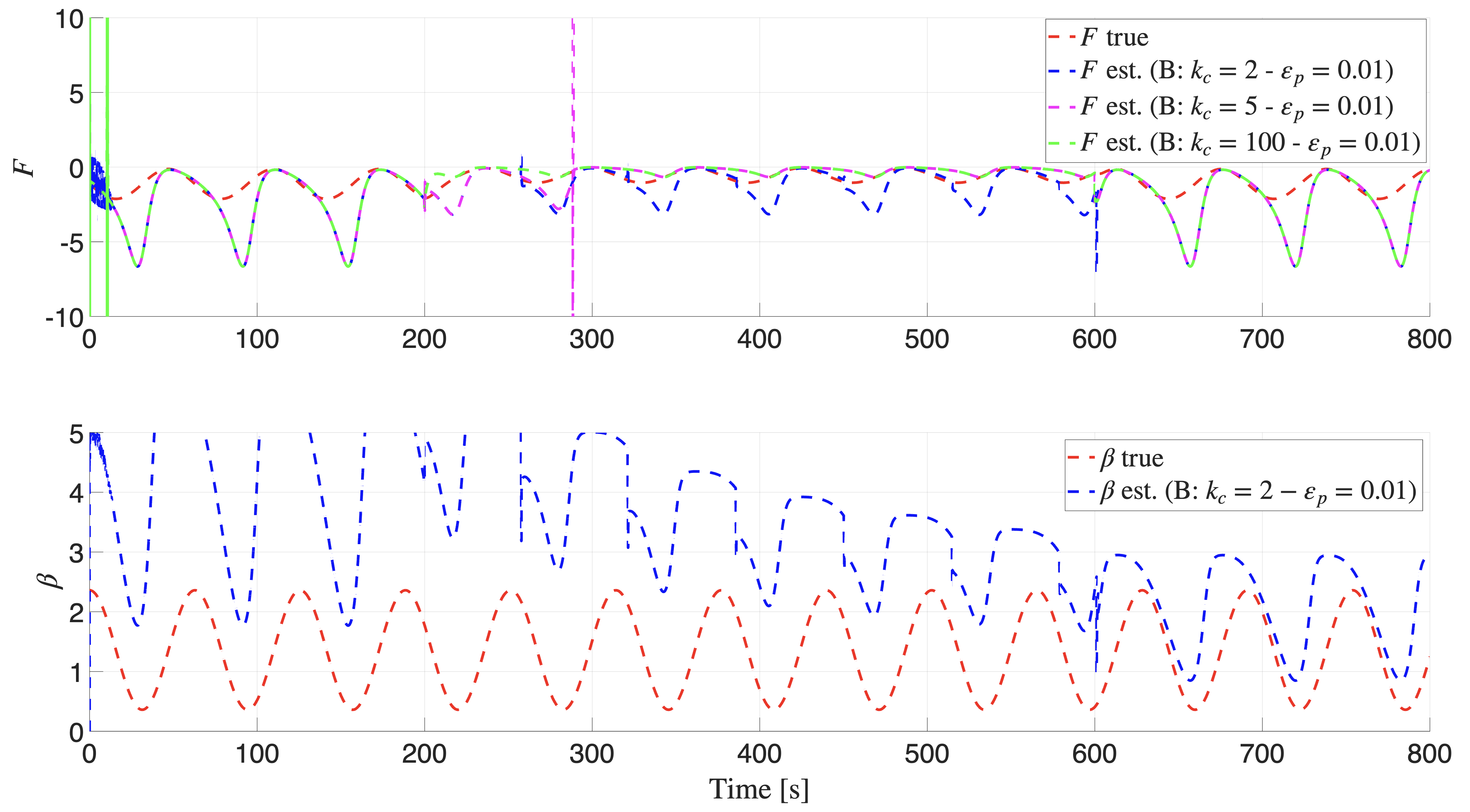}
    \caption{Method 1}
    \label{sweeping_kc_F_a}
  \end{subfigure}
  \vspace{0.5em} 
  \begin{subfigure}{\linewidth}
    \centering
    \includegraphics[width=0.9\linewidth]{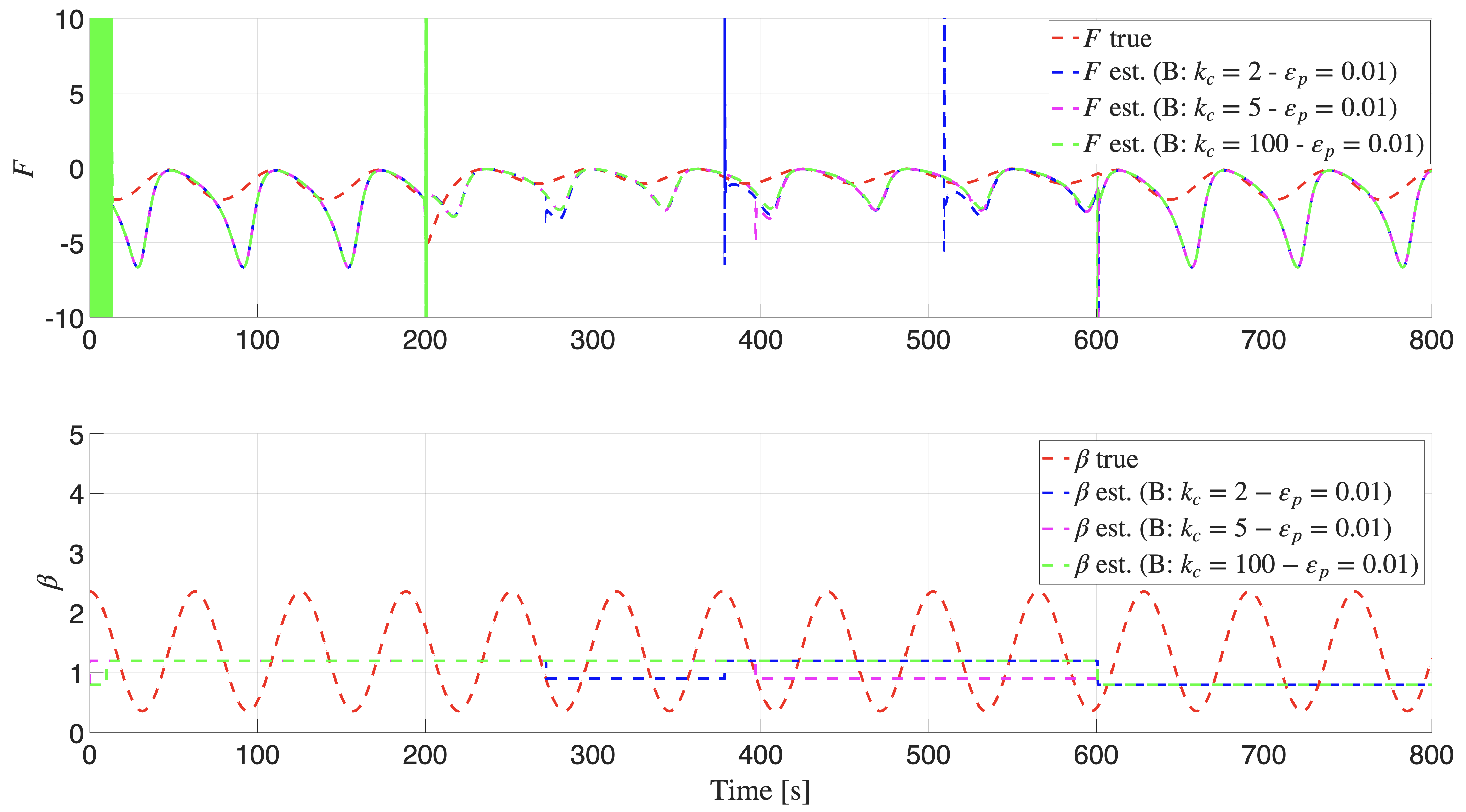}
    \caption{Method 2}  
    \label{sweeping_kc_F_b}
  \end{subfigure}
  \caption{First-order system - ($\lambda_F, \lambda_{beta}$) estimation  under $k_c$ sweeping for the method B  (using $\lambda_F = \lambda_{\beta} = 0.1$ and $\varepsilon_p = 0.01$).}
  \label{sweeping_kc_F}
\end{figure}

Under Strategy 1 (continuous gradient adaptation), Case B experiences a loss of safety  as stiffness scales up. For $k_c = 5.0$ and $k_c = 10.0$, the minimum transient distance plunges to negative values, which visually corresponds to the delayed trajectory shown in Fig. \ref{sweeping_kc_a}. Crucially, the discrete gridding framework (Strategy 2) provides a structural "salvage" effect: it maintains deterministic safety margins across all configurations, successfully mitigating transient behavior via instantaneous model selection, as illustrated in Fig. \ref{sweeping_kc_b}.
Case C, configured under a variance threshold ($\kappa = 1.0$), achieves good performance. Across both Method 1 and Method 2, and regardless of the rigidity gain ($k_c = 0.1$ up to $10.0$), the middle steady-state tracking error stays strictly centered in the neighborhood of zero. The associated time-varying profiles in Fig. \ref{sweeping_fig_1c} show an ideal, smooth asymptotic tangency condition with the safety floor $y_{\min}$, hence providing efficient parameter decoupling. Figure \ref{sweeping_kc_F} depicts the 
evolution of $F$ and $\beta$ accordingly.

        
Comparison between the two methods 
(Figs. \ref{sweeping_kc_F_a} and \ref{sweeping_kc_F_b})
highlight the fact that a higher value of $\beta$ does not satisfy necessarly safety conditions, whereas the gridding algorithm tends to converge towards lower $\beta$ value, hence
satisfying safety conditions through a diminution of the amplitude of the $F$ estimation during safety-critical interval\footnote{The scaling factor $\beta$ adapts the magnitude of the input since it is usually chosen such as $|\beta u| \sim |\dot y|$ in the case $\nu = 1$ \cite{Fliess2013}. In this sense, the adapted value of \(\beta\) provides
an indication of the order of magnitude of the term $g(x)$ in the considered system \eqref{eq:system}.}.

The comparative analysis highlights a trade-off between computational simplicity and estimation consistency when using a single 
ultra-local model versus the proposed $\beta$-gridding strategy. This leads that the strategy 2 offers simplest tuning and apparently more robust safe control with respect to the anticipation of the CBF violation.

\paragraph{Parametric study $\epsilon_p$ } 

\noindent
In this study, the stiffness of the CBF is fixed at a nominal equilibrium ($k_c = 5.0$) while the geometric penalization layer is squeezed through $\epsilon_p \in \{10^{-1}, 10^{-2}, 10^{-3}\}$. The results are presented in Tab. \ref{tab_2}.
        \begin{table}[h!]
\centering
\caption{Parametric $\epsilon_p$ sweep at fixed stiffness $k_c = 5.0$ (Nominal  effort $\Delta u_{\text{nom}} = 260.74$).}
\label{tab:eps_sweep_complete}
\fontfamily{ptm}\selectfont
\scriptsize
\setlength{\tabcolsep}{2.5pt}
\begin{tabular}{cc r ccc}
\toprule
\textbf{Method} & \textbf{Case} & \multicolumn{1}{c}{$epsilon_p$} & $\min(y)$ & ${Avg} {y}$ & $\Delta u$ \\
\midrule
& \textbf{A} & 0.100 & +0.0056 & +0.0165 & 168.97 \\
& & 0.010 & +0.0136 & +0.0276 & 168.97 \\
\textbf{Method 1} & & 0.001 & +0.0159 & +0.0305 & 168.97 \\
\cline{2-6}
\textbf{(No Grid)} & \textbf{B} & 0.100 & +0.0064 & +0.0167 & 168.97 \\
& & 0.010 & \textbf{-0.3025} & \textbf{-0.2204} & 168.97 \\
& & 0.001 & \textbf{-0.3045} & \textbf{-0.2241} & 168.97 \\
\cline{2-6}
& \textbf{C} & All & \textbf{-0.0001} & \textbf{+0.0016} & \textbf{168.97} \\
\midrule
\midrule
& \textbf{A} & 0.100 & +0.0026 & +0.0173 & 109.83 \\
& & 0.010 & +0.0099 & +0.0251 & 109.83 \\
\textbf{Method 2} & & 0.001 & +0.0136 & +0.0280 & 109.83 \\
\cline{2-6}
\textbf{(Gridding)} & \textbf{B} & 0.100 & \textbf{+0.0039} & \textbf{+0.0175} & \textbf{109.83} \\
& & 0.010 & \textbf{+0.0129} & \textbf{+0.0372} & \textbf{109.83} \\
& & 0.001 & \textbf{+0.0157} & \textbf{+0.0649} & \textbf{109.83} \\
\cline{2-6}
& \textbf{C} & All & \textbf{-0.0000} & \textbf{+0.0015} & \textbf{109.83} \\
\bottomrule
\end{tabular}
\label{tab_2}
\end{table}

\begin{figure}[pos=t]
         \centering
         \includegraphics[width=0.5\textwidth]{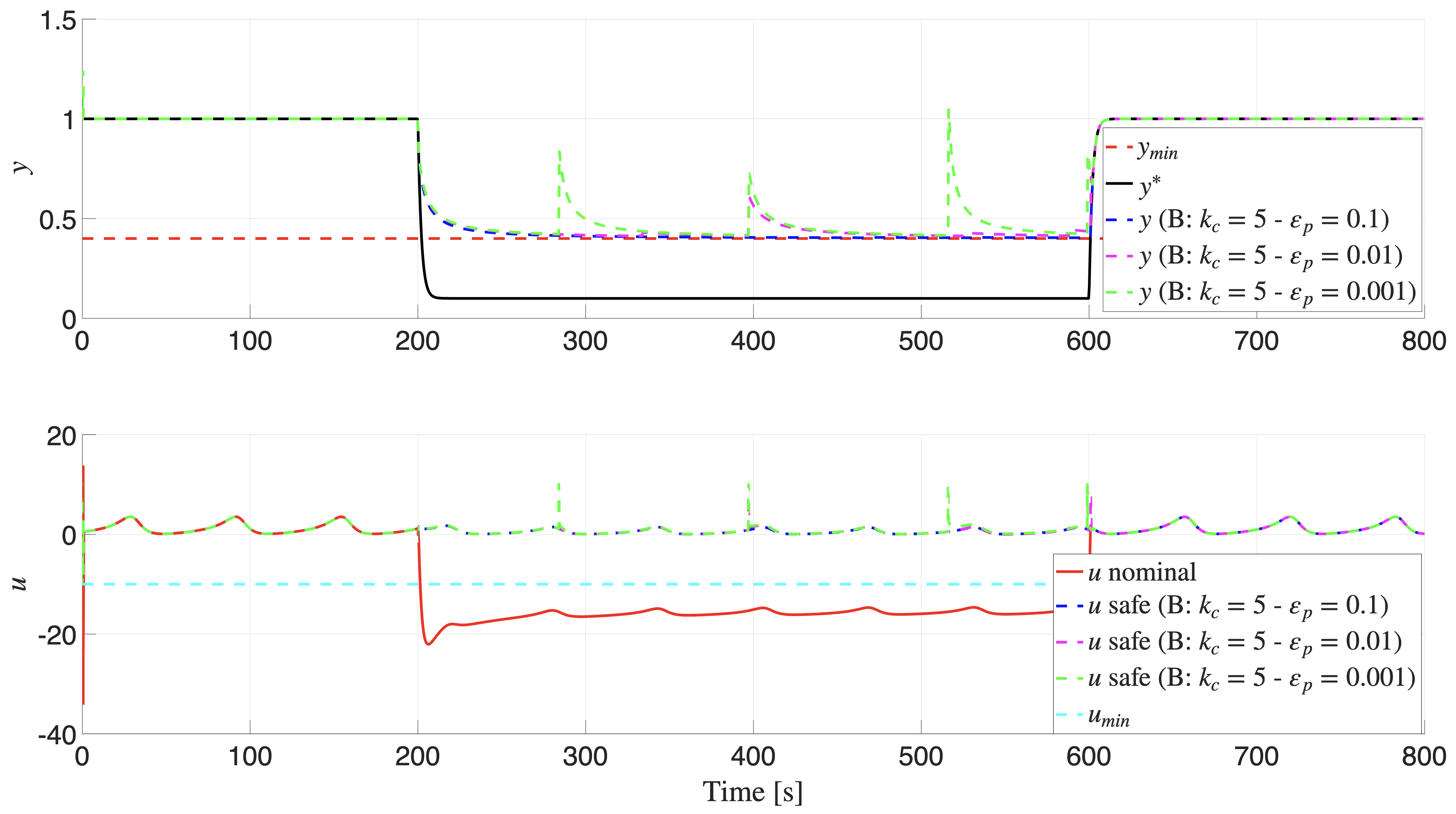}
         \caption{Method 2 for a first-order system - reference tracking under $\epsilon_p$ sweeping for the method B  (using $\lambda_F = \lambda_{\beta} = 0.1$).}
        \label{sweeping_fig_2}
        \end{figure}

The robustness of the proposed strategies when squeezing the penalty thickness parameter $\epsilon_p \in \{10^{-1}, 10^{-2}, 10^{-3}\}$ is evaluated under a fixed stiffness gain $k_c = 5.0$. 
For Method 1 (continuous gradient adaptation), Case B exhibits a severe loss of stability when $\epsilon_p \le 0.01$, resulting in  constraint violations.
In contrast, Method 2 successfully restores safety across all configurations ($\min(y) > 0$) by instantaneously switching the operational grid during sharp step transients, see Fig. \ref{sweeping_fig_2}. Furthermore, shrinking $\epsilon_p$ under Method 2 enables the trajectory to safely exploit the state space closer to the boundary, improving performance without triggering instability. 
Finally, Case C demonstrates complete structural immunity to variations in $\epsilon_p$: both Method 1 and Method 2 maintain invariant performance metrics, confirming that replacing the spatial geometric layer with the sole statistical Welford envelope ($|\mu_k| + |\sigma_k|$) effectively decouples constraint enforcement from empirical tuning parameters.

\subsection{CBF of order 2 - Active Cruise Control case}

To illustrate our proposed approach, different operating scenarios have been used within 
the active cruise control framework.
We consider the ego vehicle (E) following a preceding vehicle (PV).  
Let $x_1 = \Delta p = p_{\mathrm{PV}} - p_{\mathrm{E}}$ and 
$x_2 = \Delta v = v_{\mathrm{PV}} - v_{\mathrm{E}}$.
The state vector is \( x = [x_1,\, x_2]^\top \).

\noindent
The ego longitudinal dynamics (normalized form) are 
\begin{equation}
\left\{  \begin{aligned}
\dot x_1 &= x_2,\\
\dot x_2 &= F_{\text{g}}(x) + G\,u + {w}, 
\end{aligned}
\label{eq:system_ACC} \right.
\end{equation}
where
\[
F_{\text{g}}(x) = \frac{F_{\text{aero}}(v_E) + F_{\text{roll}}}{m} + w,
\qquad
G = -\frac{F_{\max}}{m}.
\]
and the forces are described by $F_{\mathrm{aero}}(v_E) = k_a v_E^2, \, F_{\mathrm{roll}} = k_r$ with 
$k_r = m g C_{\mathrm{roll}}$ and $k_a = \tfrac{1}{2}\rho C_d A$.

\noindent
Here, $u \in[-1,1]$ is the normalized traction/braking force, $m$ the vehicle mass, and $w$ an additive disturbance\footnote{The numerical values of each parameter are $m$ = 1500 {kg},
$\rho$ = 1.30 kg/m$^3$, $C_d$ = 0.32, $A$ = 2.4 
m$^2$, $g$ = 9.81 m/s$^2$, $C_{\mathrm{roll}}$ = 0.01, $F_{\max}$ = $0.3 m g$ = 4414.5 N.
}. The inter-vehicle distance $x_1$ is of relative degree of order 2.

{The control problem is to track the inter-vehicle distance $x_1$, following a reference trajectory $x_1^*$ and ensuring the safety constraint.
The tracking is made using a modified version of the model-free control approach that has been proposed in \cite{michel2018paramodel} by the first author\footnote{See e.g. recent applications in HIV epidemilogy \cite{michelHIV}, 
CFD-based Navier-Stokes \cite{ECC_ISIS} and wind turbines control \cite{michelOpenFAST}.}.}

\subsubsection*{Numerical results}

The proposed $\beta$-gridding strategy, described in the Algorithm \eqref{Algo1}, operates at the controller sampling period \(T_s\) (chosen at 20~ms).  
All quantities \(x_1,\dot x_1,\ddot x_1\) are computed in discrete time.  

The following CBF parameters have been chosen to damp oscillations
$k_1 = 2.0, k_2 = 3.0,T_s = 0.02\,\mathrm{s}, d_{\min} = 2\,\mathrm{m}, t_h = 0.5\,\mathrm{s}$ and method C has been used since it offers the best robustness towards the tangency condition with the safety minimal distance. Following the formulation of \cite{Castro2022}, the tracking is expressed as
$x_1^* = d_0 + t_h v_E$ where $v_E$ is the ego-vehicle velocity. In the nominal conditions, the PV has a constant speed of 15 m/s. The initial conditions are set to $x_1(0) = 10$ m and $x_2(0) = 2$ m/s. 


\noindent

\begin{figure}[pos=t]
         \centering
         \includegraphics[width=0.5\textwidth]{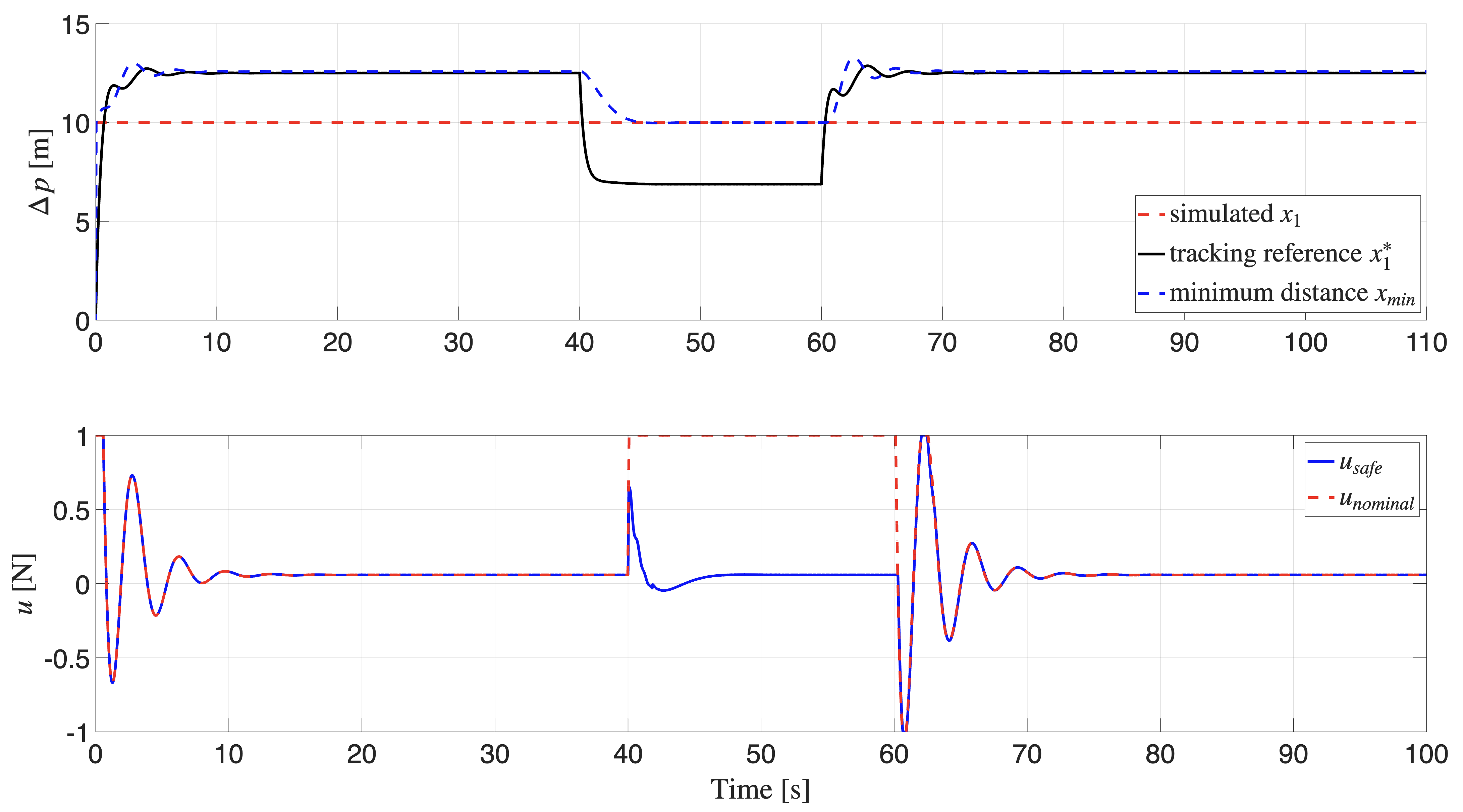}
         \caption{ACC case - Tracking of $\Delta p$ including a temporarily unsafe modification of the reference (between 40 and 60 s).}
        \label{fig:ACC_fig_1}
        \end{figure}

\begin{figure}[pos=t]
         \centering
         \includegraphics[width=0.5\textwidth]{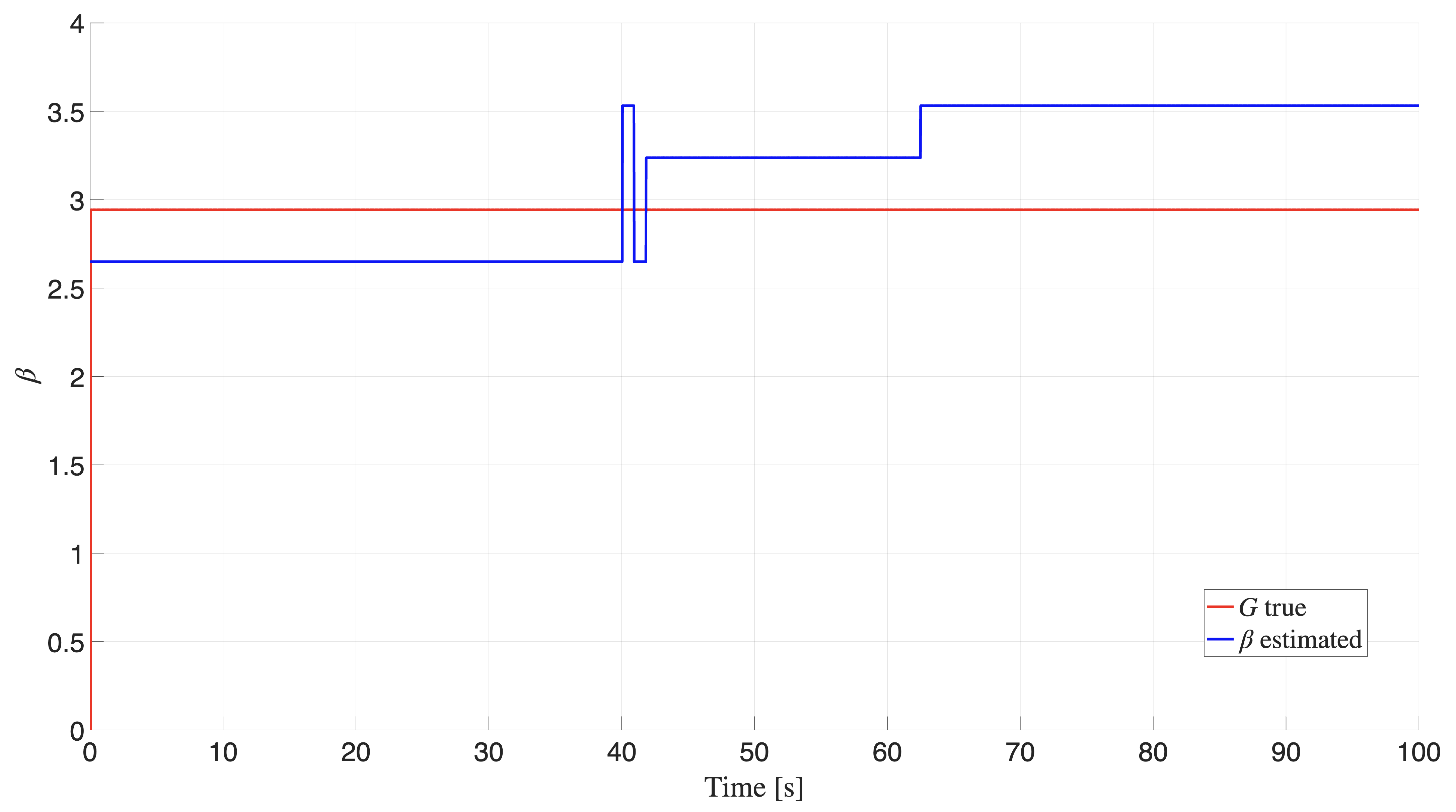}
         \caption{ACC case - Evolution of $\beta$ through the gridding algorithm.}
        \label{fig:ACC_fig_2}
        \end{figure}

To highlight the role of the CBF safety filter, a scenario is considered where the tracking reference 
temporarily violates the minimum safety distance. In this situation, the nominal controller alone would 
drive the system toward an unsafe configuration. 
The proposed CBF-based controller automatically overrides the nominal action and maintains the inter-vehicle 
distance above the prescribed safety threshold.
Figure \ref{fig:ACC_fig_1} illustrates the operation of the active cruise control considering the tracking of the inter-vehicle distance subjected to a fall-down of the reference below the safety minimum distance and Fig. \ref{fig:ACC_fig_2} shows the evolution of $\beta$, which is driven by the Algorithm \ref{Algo1}.

At the beginning of the simulation, while the control is being initialized, the CBF computation is not enabled. The gridding algorithm is triggered once the control is stabilized (from 10 s).  
This experiment highlights that the proposed framework can enforce safety constraints even when the nominal reference becomes temporarily infeasible, without relying on a precise vehicle model. This illustrates the relevance of the proposed approach for safety-critical automotive applications where model uncertainty and inconsistent references may occur in practice.
Compared with the method 1, the gridding strategy improves robustness of the safety filter by selecting the most suitable input–output scaling according to the current operating regime, which is particularly important in ACC scenarios where aerodynamic forces and rolling resistance vary with velocity.

\section{Conclusion and perspectives}

This work introduced a computational framework for the construction of high-order Control Barrier Functions (CBFs) within a model-free setting, relying on an ultra-local modeling approach that is continuously updated online.
The proposed method combines a grid-based strategy with an ultra-local model-based CBF formulation, enabling the systematic exploration of parameter sets and the selection of optimal configurations to refine the online estimation of the unknown system dynamics. This adaptive mechanism enhances robustness with respect to disturbances and model uncertainties.

This work opens several promising research directions. A key perspective lies in the derivation of formal safety guarantees by explicitly accounting for the estimation error induced by the ultra-local model. 
In addition, the proposed approach is particularly well suited for applications involving complex and poorly known dynamics, such as battery charging systems. In this context, safety-critical constraints related to voltage, current, and temperature could be enforced without relying on detailed electrochemical models, paving the way for adaptive and safe fast-charging strategies.

\section*{Declaration of generative AI and AI-assisted technologies in the manuscript preparation process}

During the preparation of this work, the authors used ChatGPT (OpenAI) and Gemini (Google) in order to assist with manuscript organization, English language editing, and code refinement. After using these tools, the authors reviewed and edited the content as needed and take full responsibility for the content of the published article.

\bibliographystyle{cas-model2-names}






\end{document}

%% file: Figures/Architecture_scheme_v1.tex
\begin{figure}[pos=t]
\centering
\resizebox{0.5\textwidth}{!}{
\begin{tikzpicture}[
block/.style={
    draw, rectangle,
    text width=3.4cm,
    minimum height=1.2cm,
    align=center,
    fill=syscolor!25
},
smallblock/.style={
    draw, rectangle,
    text width=2.4cm,
    minimum height=1.2cm,
    align=center,
    fill=syscolor!25
},
cbfsmall/.style={
    draw, rectangle,
    text width=2.8cm,
    minimum height=1.2cm,
    align=center,
    fill=cbfcolor!35
},
estblock/.style={
    draw, rectangle,
    text width=4cm,
    align=center,
    fill=gpcolor!30
},
groupbox/.style={
    draw,
    rectangle,
    rounded corners,
    dashed,
    inner sep=10pt
},
>=Latex
]

\node[block] (controller) at (-4,0) {Nominal Controller};
\node[cbfsmall] (cbf) at (0,0) {CBF ($\hat{\Delta}_m$ est.)};
\node[smallblock] (plant) at (3.5,0) {Plant};

\node[estblock] (delta) at (-2.5,-3) {
\textbf{Method 1} \\[2pt]
$(F,\beta)$ gradient descent 
};

\node[estblock] (model) at (2.5,-3) {
\textbf{Method 2} \\[2pt]
$\beta$ gridding + decision 
};

\node[groupbox, fit=(delta)(model), label=below:{{Online Estimation}}] (group) {};

\draw[->] (controller) -- node[above] {$u_n$} (cbf);
\draw[->] (cbf) -- node[above] {$u$} (plant);

\draw[->] (plant.south) -- ++(0,-1.2)
           -- ++(-7.5,0)
           -- (controller.south);

\node at (4,-1.2) [right] {$y$};

\draw[->] (delta.north) -- ([xshift=-6mm]cbf.south);
\draw[->] (model.north) -- ([xshift=6mm]cbf.south);

\end{tikzpicture}
}
\caption{Proposed CBF architecture including online $\hat{\Delta}_m$ and ultra-local model updates.}
\label{archi_CBF}
\end{figure}

%% file: Figures/Welford_scheme_v1.tex
\begin{figure}[pos=t]
\centering
\resizebox{0.5\textwidth}{!}{
\begin{tikzpicture}[scale=1.0,>=stealth]


\draw[->] (2,-5) -- (12,-5);

\node at (2.5,-5.5) {$k-N$};

\node at (2.5,-1.5) {$\hat{F}$};

\node at (11.5,-5.5) {$k$};

\foreach \x/\y in {
3/-3,
4/-4,
5/-3.6,
6/-4.8,
7/-3.7,
8/-2.2,
9/-4.2,
10/-2.4,
11/-2.6}
{
\filldraw[blue!60!black] (\x,\y) circle (2pt);
}

\draw[dashed, purple] (2.5,-3.5) -- (11.5,-3.5);
\node[right,purple] at (11.5,-3.5)
{$\mu_k^{\Delta m}$};

\draw[dashed, red] (2.5,-2.7) -- (11.5,-2.7);
\draw[dashed, red] (2.5,-4.3) -- (11.5,-4.3);

\node[right,red] at (11.5,-2.7)
{$\sigma_k^{\Delta m}$};

\draw[dashed, orange] (2.5,-2.2) rectangle (11.5,-4.8);

\node[right,orange] at (11.5,-2.2)
{$\max \Delta m_k$};

\node[right,orange] at (11.5,-4.8)
{$\min \Delta m_k$};

\draw[black][->] (2.5,-5.2) -- (2.5,-2);

\draw[black] (11.5,-5.2) -- (11.5,-2);

\end{tikzpicture}
}
\caption{Construction of the uncertainty envelope $\Delta m_k$ based on the online Welford estimation of $(\mu_k, \, \sigma_k)$ from $\hat{F}_k$ and $\max_{[k-N, \, k]}|\Delta m_k|$ over a sliding estimation window.}
\label{Welford_scheme}
\end{figure}

%% file: Figures/ULM_Delta_m.tex
\begin{figure}[pos=t]
\centering
\begin{tikzpicture}[scale=1]

\node at (0,4) {${F}$};

\draw[->] (-0.2,0) -- (7,0) node[right] {$t$};
\draw[->] (0,-0.2) -- (0,3.5);

\draw[dashed] (1.5,0) -- (1.5,3);
\node[below] at (1.5,0) {$t_k$};

\draw[dashed] (5.5,0) -- (5.5,3);
\node[below] at (5.5,0) {$t_{k+1}$};

\draw[red, thick]
plot[smooth] coordinates
{(0.5,0.6) (1.0,0.8) (1.5,1.0) (2.2,1.3) (3,1.6) (4,2.0) (5.6,2.5)}
node[right] {$\hat F(t)$};

\draw[black, thick]
plot[smooth] coordinates
{(0.5,0.7) (1.0,0.6) (1.5,1.1) (2.0,0.9) (2.7,1.5) (3.3,1.4)
(4.0,2.1) (4.6,1.9) (5.2,2.4) (5.6,2.5)};

\node at (6.1,2.1) {$F(t)$};

\draw[dashed, blue, thick]
plot[smooth] coordinates
{(1.5,1.3) (2.2,1.7) (3,2.1) (4,2.6) (5.5,3.1)};

\draw[dashed, blue, thick]
plot[smooth] coordinates
{(1.5,0.7) (2.2,0.9) (3,1.1) (4,1.4) (5.5,1.8)};

\node[blue] at (6.5,3.0) {$\hat F(t) + \hat{\Delta}_m$};
\node[blue] at (6.5,1.8) {$\hat F(t) - \hat{\Delta}_m$};

\draw[<->, thick] (2.5,1.0) -- (2.5,1.8);
\node[right] at (2.5,1.4) {$\Delta_m$};


\end{tikzpicture}
\caption{Uncertainty envelope associated
to the ultra-local dynamics. Starting from the current time $t_k$,
a Welford-based prediction of the estimated dynamics $\hat F(t)$ is
performed over a short horizon $h$. The bound $\Delta_m$ defines an
uncertainty envelope that accounts for estimation errors and local
variations of the underlying system dynamics.}
\label{ULM_Deltam}
\end{figure}